\documentclass[10pt,twocolumn]{article}
\usepackage{graphicx}
\usepackage{amsmath,amssymb,times,cite}
\usepackage{xcolor}
\newcommand{\vc}[3]{\overset{#2}{\underset{#3}{#1}}}
\newcommand{\op}[1]{\operatorname{#1}}
\usepackage{tabularx}
\newcommand{\bg}[1]{\boldsymbol{#1}} %Bold Greek letters
\newcommand{\bm}[1]{\mathbf{#1}} %Bold vectors and matrices
\usepackage{multirow}
\newcommand\T{{\mathpalette\raiseT\intercal}}
\newcommand\raiseT[2]{%
\setbox0\hbox{$#1{#2}$}\raise\dp0\box0}

\usepackage{array}
\newcolumntype{M}[1]{>{\centering\arraybackslash}m{#1}}

\usepackage{booktabs}
\title{\LARGE\textbf{Adaptive Spectral Graph Wavelets for Collaborative Filtering}}
\author{Osama Alshareet and A. Ben Hamza\\
Concordia Institute for Information Systems Engineering\\
Concordia University, Montreal, QC, Canada
}
\date{}

\begin{document}
\maketitle

\begin{abstract}
Collaborative filtering is a popular approach in recommender systems, whose objective is to provide personalized item suggestions to potential users based on their purchase or browsing history. However, personalized recommendations require considerable amount of behavioral data on users, which is usually unavailable for new users, giving rise to the cold-start problem. To help alleviate this challenging problem, we introduce a spectral graph wavelet collaborative filtering framework for implicit feedback data, where users, items and their interactions are represented as a bipartite graph. Specifically, we first propose an adaptive transfer function by leveraging a power transform with the goal of stabilizing the variance of graph frequencies in the spectral domain. Then, we design a deep recommendation model for efficient learning of low-dimensional embeddings of users and items using spectral graph wavelets in an end-to-end fashion. In addition to capturing the graph's local and global structures, our approach yields localization of graph signals in both spatial and spectral domains, and hence not only learns discriminative representations of users and items, but also promotes the recommendation quality. The effectiveness of our proposed model is demonstrated through extensive experiments on real-world benchmark datasets, achieving better recommendation performance compared with strong baseline methods.
\end{abstract}

\bigskip
\noindent\textbf{Keywords}:\, Collaborative filtering; recommendation; deep learning; spectral graph wavelets; Box-Cox transformation.

\section{Introduction}
Recommender systems are instrumental in improving users' web browsing experience and retaining customers by providing tailored suggestions specific to the customer's needs and helping users find items of interest~\cite{Ziou:08}. These systems provide valuable insight into consumers' purchasing behavior and browsing activity, and have proven to be effective not only at driving sales up in e-commerce, but also at increasing the number of people signing up for and using social media sites. When it comes to buying online, a convenient shopping experience that is tailored to consumers is probably the most important factor, with next day delivery pretty much standard now. Also, many consumers are turning to Internet shopping, particularly during pandemic lockdowns, opening up new opportunities for online retailers and other sellers of consumer goods.

Collaborative filtering is a widely used technique by recommender systems~\cite{Konstan:97,YHU:08}, and aims at identifying user preferences using explicit feedback such as user ratings (e.g., 5-star or 10-point user rating systems) or implicit feedback such as user interactions with items (e.g., purchase, watch, listen, or click on an item). For the former, a vast majority of approaches relies on matrix factorization, which represents users and items in a low-dimensional latent space~\cite{Koren:09}. While explicit feedback includes explicit input by users regarding their interest in items, it is, however, not always available. By contrast, implicit feedback, in which user preferences are expressed via item interactions, is more common and practical, but poses challenges for recommender systems. In collaborative filtering with implicit feedback, which is the focus of our work, the lack of interaction of a user with an item does not necessarily mean that an item is irrelevant for the user. A popular approach for collaborative filtering with implicit feedback is Bayesian personalized ranking (BPR) based matrix factorization~\cite{Rendle:09}. BPR is a learning-to-rank method that produces a personalized ranking list of recommendations using a pairwise loss function, and assumes that observed user-item interactions should be ranked higher than the unobserved ones. While matrix factorization models are simple and effective~\cite{Koren:09,YKoren:08,YKoren:10}, they are inherently linear and hence unable to capture the nonlinear structure of the implicit feedback data.

In recent years, a plethora of deep neural networks and graph-based recommendation models have been applied to the collaborative filtering setting, showing improved recommendation performance~\cite{Xiangnan:17,LZheng:18,XWang:19,DLiang:18,XiangWang:20,JianxinChang:20,ZhaopengLi:20,Smola:20,LChen:20,Zhaopeng:20}. This boost in performance, compared to linear recommendation models, is attributed in large part to the fact that deep neural networks learn nonlinear features obtained from stacking up feature extractors, which are passed through non-linear activation functions, and hence capture more complex patterns. Also, the success of deep neural networks in recommender systems and other downstream tasks has been greatly accelerated by using graphics processing units, which have become the platform of choice for training large, complex learning systems.

Previous work~\cite{LZheng:18} on graph-based recommendation via graph convolutional networks has recently demonstrated that collaborative filtering based on spectral graph theory is capable of discovering deep connections between users and items, and hence helps alleviate the cold-start problem. The general idea in spectral graph theory is that to any graph we may associate a corresponding matrix, which records information about its structure through the matrix spectrum. While the spectral collaborative filtering model provides good recommendation performance~\cite{LZheng:18}, its convolution operation is, however, defined in terms of the graph Fourier bases, which yield localization only in the spectral domain. In this paper, we propose a deep recommendation model, called CAGLE, for \textbf{C}ollaborative Filtering with \textbf{A}daptive Spectral \textbf{G}raph Wave\textbf{LE}ts on graph-structured data. More precisely, we introduce an adaptive graph wavelet basis that captures the graph's global structure and yields localization of graph signals in both spatial and spectral domains. The proposed collaborative filtering approach leverages graph neural networks in the spectral graph theoretic setting, and employs a power transform to stabilize the variance of the graph frequencies, which are attenuated while learning the embeddings of users and items. The main contributions of this work can be summarized as follows:
\begin{itemize}
\item We introduce an adaptive graph wavelet basis by leveraging the power transformed eigenvalues of the graph Laplacian in conjunction with spectral graph wavelets, yielding localization of graph signals in both spatial and spectral domains.
\item We develop a deep recommendation model (CAGLE) for efficiently learning low-dimensional embeddings of users and items in a bipartite graph by leveraging the properties of spectral graph wavelets.
\item We demonstrate through extensive experiments the superior performance of CAGLE over state-of-the-art baselines on several benchmark datasets.
\end{itemize}	

\medskip The rest of this paper is organized as follows. In Section 2, we review important relevant work. In Section 3, we introduce a deep recommendation model using adaptive spectral graph wavelets. We discuss in detail the main components of the proposed collaborative filtering framework for implicit data. In Section 4, we present experimental results to demonstrate the competitive performance of our recommendation model on real-world benchmark datasets. Finally, we conclude in Section 5 and point out future work directions.

\section{Related Work}
The basic goal of collaborative filtering is to generate personalized recommendations by leveraging historical data about interactions between users and items. To achieve this goal, a variety of collaborative filtering approaches have been proposed for learning with implicit feedback.

\medskip\noindent\textbf{Deep Neural Networks for Recommendation.}\quad Much of the recent work in recommender systems leverages deep learning~\cite{Sedhain:15,HaoWang:15,Xiangnan:17,DLiang:18,Hui2022ReBKC}, which has shown remarkable capabilities in learning discriminative feature representations by extracting high-level features from data using multilayered neural networks. He \textit{et al.}~\cite{Xiangnan:17} introduce neural collaborative filtering (NCF), a deep neural network that integrates the linear matrix factorization with the non-linear multi-layer perceptron in an effort to learn the interaction function that maps model parameters to the predicted interaction score through nonlinear neural optimization in lieu of the inner product of latent factors. A major drawback of NCF is that the number of the model's parameters grows linearly with both the number of users and items. Liang \textit{et al.}~\cite{DLiang:18} extend variational autoencoders to collaborative filtering for implicit feedback using a generative model with a multinomial conditional likelihood function parameterized by a neural network. To learn the parameters of this generative model, the posterior distribution is approximated using variational inference. However, the choice of the prior may negatively impact the performance, specially when a lot of data is available for a user.

\medskip\noindent\textbf{Graph Convolutional Networks for Recommendation.}\quad More recently, graph convolutional networks (GCNs)~\cite{Kipf:17}, which are an efficient variant of convolutional neural networks on graph-structured data, have proven to be useful in many graph analysis tasks, achieving state-of-the-art performance in various application domains, including recommendation~\cite{LZheng:18,XWang:19,McAuley:20,XiangnanHe:20,yu2022graph,xia2022hypergraph,liu2020deoscillated,Song2022DBRGA,Qian2021DAGNN,Qian2023PIGNN,Fan2019GraphRec,Wu2021SGL}. Zheng \textit{et al.}~\cite{LZheng:18} propose spectral collaborative filtering (SpectralCF), a deep recommendation model based on spectral filtering of graph signals using the graph Fourier transform. SpectralCF formulates the relationships between users and items as a bipartite graph, and employs polynomial approximations of the spectral filters in order to help alleviate the learning complexity of the model~\cite{Defferrard:16}. Wang \textit{et al.}~\cite{XWang:19} introduce neural graph collaborative filtering (NGCF), a GCN-based recommendation model that directly encodes the collaborative information of users by leveraging the high-order connectivity from user-item interactions via embedding propagation. A higher-order GCN-based approach for collaborative filtering is also presented in~\cite{McAuley:20}, where multiple stacked mixed-order GCN layers are used, followed by an average pooling layer for information fusion. He \textit{et al.}~\cite{XiangnanHe:20} present a simplified NGCF model by leveraging the simple graph convolution~\cite{Wu:19}, which successively removes the nonlinear activation functions and collapses the weight matrices between consecutive layers. In this simplified model, the low-dimensional embeddings of users and items are learned iteratively for a pre-defined number of power iteration steps, and then combined using a weighted average to obtain the final representations. Our proposed framework differs from existing collaborative filtering approaches in two main aspects. First, we design an adaptive filter by leveraging a power transform with the aim to stabilize the variance of graph frequencies in the spectral domain. Second, we develop a deep recommendation model that captures not only the graph's global structure, but also yields localization of graph signals in both spatial and spectral domains.

\section{Method}
\begin{figure}[!htb]
\centering
\includegraphics[scale=.5]{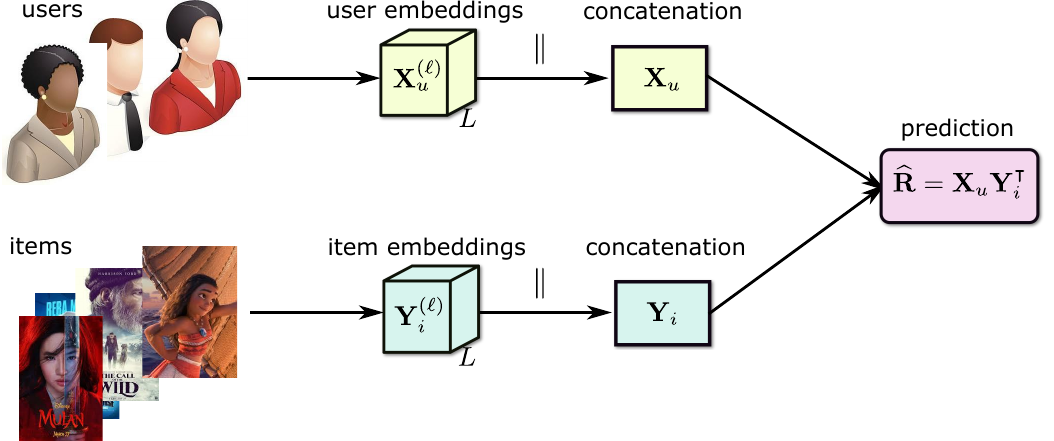}
\caption{Schematic diagram of the proposed framework.}
\label{fig:CAGLE}
\end{figure}

In this section, we first introduce our notation and provide a brief background on spectral collaborative filtering, followed by the problem formulation of collaborative filtering with implicit feedback on bipartite graphs. A bipartite graph is a graph whose nodes can be partitioned into two disjoint sets such that there is no edge that connects nodes from the same set~\cite{Godsil:01}. Then, we present the main building blocks of our proposed graph wavelet collaborative filtering framework, which is illustrated in Figure~\ref{fig:CAGLE}. Each layer of our model takes the initialized user/item embeddings as input and returns embeddings that are obtained by applying a layer-wise propagation rule. Then, the final user/items embeddings are acquired by concatenating the representations learned at different layers. Finally, we take the inner product to predict the preference of a user toward an item.

\subsection{Preliminaries and Problem Statement}
\begin{figure}[!htb]
\centering
\includegraphics[scale=0.64]{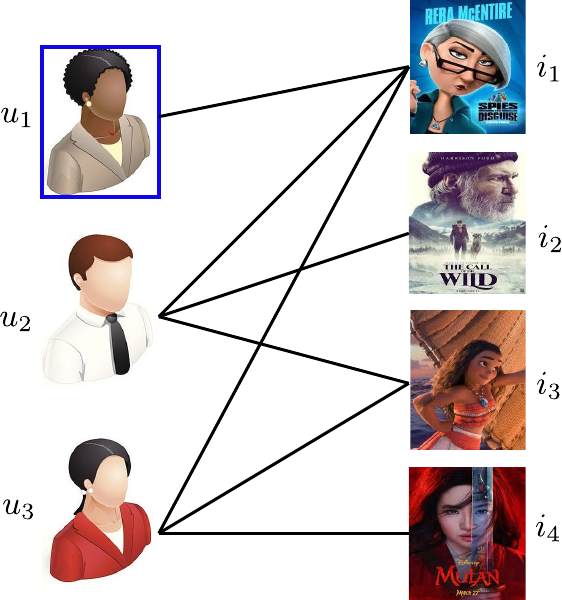}
\caption{Illustration of a user-item bipartite graph.}
\label{fig:bipartite}
\end{figure}

A bipartite user-item graph $\mathcal{G}=(\mathcal{U},\mathcal{I},\mathcal{E})$ consists of a node set $\mathcal{U}=\{u_1,\dots,u_M\}$ of users, a node set $\mathcal{I}=\{i_1,\dots,i_K\}$ of items, and an edge set $\mathcal{E}$. An edge $e=(u,i)\in \mathcal{E}$ indicates that item $i$ is of interest to user $u$. The node sets $\mathcal{U}$ and $\mathcal{I}$ are disjoints, and their union $\mathcal{U}\cup\mathcal{I}$ consists of $N=M+K$ nodes. It is worth mentioning that a bipartite graph is 2-colorable (i.e., each node can be assigned one of two colors) and has no odd length cycles (i.e., no cycles with odd number of edges). An illustrative example of a bipartite graph with 3 users and 4 items is depicted in Figure~\ref{fig:bipartite}, which shows that item $i_3$ is more likely to be of interest to the cold-start (target) user $u_1$, as there exist two paths connecting them compared to a single path between $u_1$ and the other items.

\medskip\noindent\textbf{Adjacency Matrix:}\quad The adjacency matrix of the bipartite user-item graph is defined as
\begin{equation}
\bm{A}=\begin{bmatrix}
\bm{0} & \bm{R}\\
\bm{R}^{\T} & \bm{0}
\end{bmatrix},
\end{equation}
where $\bm{R}=(\bm{R}_{ui})$ is an $M\times K$ implicit feedback matrix (also called biadjacency or user-item interaction matrix) representing the users' binary feedback (relevancy) for the items such that $\bm{R}_{ui}=1$ if item $i$ is an implicit interaction between user $u$ and item $i$ is observed; and 0 otherwise. Note that $\bm{R}_{ui}=1$ does not necessarily mean that $u$ likes $i$, just that the user may certainly have interest in the item. In a similar vein, $\bm{R}_{ui}=0$ does not necessarily mean that $u$ dislikes $i$, just that the user may be unaware of the item.

\medskip\noindent\textbf{Normalized Laplacian Matrix:}\quad The normalized Laplacian matrix of the bipartite user-item graph is defined as
\begin{equation}
\bm{L}=\bm{I}-\bm{D}^{-\frac{1}{2}}\bm{A}\bm{D}^{-\frac{1}{2}},
\end{equation}
where $\bm{D}=\op{diag}(\bm{A}\bm{1})$ is the diagonal degree matrix, and $\bm{1}$ is an $N$-dimensional vector of all ones.

\medskip\noindent\textbf{Graph Fourier Basis:}\quad Since the normalized Laplacian matrix is symmetric positive semi-definite~\cite{Krim:15}, it admits an eigendecomposition given by $\bm{L}=\bm{\Phi}\bm{\Lambda}\bm{\Phi}^{\T}$, where $\bm{\Phi}=(\bg{\varphi}_1,\dots,\bg{\varphi}_N)$ is an orthonormal matrix whose columns constitute an orthonormal basis of eigenvectors (called Graph Fourier basis) and $\bm{\Lambda}=\op{diag}(\lambda_1,\dots,\lambda_N)$ is a diagonal matrix comprised of the corresponding eigenvalues such that $0=\lambda_1\le\dots\le\lambda_N= 2$. Note that $\lambda_N= 2$ only holds for bipartite graphs.

\medskip\noindent\textbf{Graph Wavelet Basis:}\quad Spectral graph wavelets have shown to allow localization of graph signals in both spatial
and spectral domains~\cite{Hammond:11,Chunyuan:13,Claire:18,Bingbing:19}. Let $g_{s}(\lambda)=e^{-\lambda s}$ be the transfer function (also called frequency response) of the heat kernel with scaling parameter $s$. The spectral graph wavelet basis $\bm{\Psi}_{s}$ is defined as
\begin{equation}
\bm{\Psi}_{s}=\bm{\Phi}\bm{G}_{s}\bm{\Phi}^{\T},
\end{equation}
where $\bm{G}_{s}=g_{s}(\bm{\Lambda})=\op{diag}(g_{s}(\lambda_1),\dots,g_{s}(\lambda_N))$ is a diagonal matrix. Note that $\bm{\Psi}_{s}$ is also referred to as the heat kernel matrix whose inverse $\bm{\Psi}_{s}^{-1}$ is obtained by simply replacing the scale parameter $s$ with its negative value.

\medskip\noindent\textbf{Spectral Collaborative Filtering:}\quad Given input user embedding matrix $\bm{X}_{u}^{(\ell)}\in\mathbb{R}^{M\times F_{\ell}}$ and input item embedding matrix $\bm{Y}_{i}^{(\ell)}\in\mathbb{R}^{K\times F_{\ell}}$ of the $\ell$-th layer with $F_{\ell}$ feature maps, the output feature matrices $\bm{X}_{u}^{(\ell+1)}$ and $\bm{Y}_{i}^{(\ell+1)}$ of spectral collaborative filtering (SpectralCF) are obtained by applying the following layer-wise propagation rule~\cite{LZheng:18}:
\begin{equation}
\begin{bmatrix}
\bm{X}_{u}^{(\ell+1)} \\
\bm{Y}_{i}^{(\ell+1)}
\end{bmatrix}=\sigma\left(\bm{\Phi}\widetilde{\bm{\Lambda}}\bm{\Phi}^{\T}
\begin{bmatrix}
\bm{X}_{u}^{(\ell)} \\
\bm{Y}_{i}^{(\ell)}
\end{bmatrix}\bm{W}^{(\ell)}\right),
\label{Eq:AGCNprop}
\end{equation}
for $\ell=0,\dots,L-1$, where $\bm{W}^{(\ell)}\in\mathbb{R}^{F_{\ell}\times F_{\ell+1}}$ is a trainable weight matrix  with $F_{\ell +1}$ feature maps, $\sigma(\cdot)$ is the point-wise sigmoid activation function, and $\widetilde{\bm{\Lambda}}=\bm{I}+\bm{\Lambda}$ is a diagonal matrix whose diagonal elements $\tilde{\lambda}_{i}$ are all positive.

\medskip\noindent\textbf{Problem Statement.}\quad The objective of collaborative filtering for implicit feedback is to leverage the user-item interaction matrix to estimate user-item preference scores for unobserved interactions. Denoting by $\mathcal{I}_{u}^{+}$ the set of items that user $u$ has previously interacted with and $\mathcal{I}_{u}^{-}=\mathcal{I}\setminus\mathcal{I}_{u}^{+}$ its complement set, the goal of item recommendation with implicit feedback is to generate a top-K ranking of items from $\mathcal{I}_{u}^{-}$ that user $u$ is most likely to prefer (e.g., purchase, watch, listen, or like).

\subsection{Proposed Method}
\noindent\textbf{Motivation.}\quad Spectral graph wavelet bases can yield localization of graph signals in both spatial and spectral domains whereas graph Fourier bases yield localization only in the spectral domain. In addition, graph wavelet bases are sparser than their graph Fourier counterparts, making them more computationally efficient. Unlike the graph Fourier basis, the graph wavelet basis is multiscale and hence can capture both local and global information from different graph neighborhoods by varying the value of the scaling parameter, which controls the amount of diffusion on the graph. For small values of the scaling parameter, the graph wavelet basis is determined by small (immediate) neighborhoods of a given node, reflecting local properties of the graph around that node. The larger the scaling parameter, the bigger is the support of the graph wavelet basis on the graph and hence the wavelet basis captures the graph's global structure.

\medskip\noindent\textbf{Box}-\textbf{Cox Transformation.}\quad The spectrum of the normalized Laplacian matrix contains the structural information of the graph, as the multiplicity of the zero eigenvalue is exactly the number of connected components of a graph, while the second smallest eigenvalue is a global property that generally quantifies how well connected the graph is. In addition, the largest eigenvalue has an intuitive interpretation in the sense that how close is a graph is to being bipartite.

Each eigenpair $(\lambda_i,\bg{\varphi}_i)$ of the normalized Laplacian matrix can be interpreted in a similar fashion as the classical Fourier basis: the eigenvalues $\lambda_i$ act as frequencies, while the eigenvectors $\bg{\varphi}_i$ play the role of the
complex exponentials (sinusoidals). Since $\lambda_i=\bg{\varphi}_{i}^{\T}\bm{L}\bg{\varphi}_i$, we can interpret each eigenvalue as a measure of variation of its corresponding eigenvector. Eigenvectors associated with larger eigenvalues oscillate more rapidly, similar to the behaviors of classical Fourier basis functions in the Euclidean domain.

In order to stabilize the eigenvalues' variance, we can use a power transform~\cite{Yeo:00}, which essentially transforms non-normally distributed data to a set of data that has approximately normal distribution. More specifically, since the elements of $\widetilde{\bm{\Lambda}}$ are positive (i.e., $\tilde{\lambda}_{i}>0$), we apply the Cox-Box transformation given by
\begin{equation}
\tilde{\lambda}_{i}^{(\kappa)} = \begin{cases}
(\tilde{\lambda}_{i}^{\kappa}-1)/\kappa &\mbox{if } \kappa \neq 0 \\
\log\tilde{\lambda}_{i} & \mbox{if } \kappa = 0
\end{cases}
\label{Eq:BoxCox}
\end{equation}
where $\kappa$ is the transformation parameter, which is estimated by maximizing the log-likelihood function using the boxcox function in SciPy, an open-source Python library for scientific and statistical computing. More specifically, the SciPy boxcox function takes as input the elements of $\widetilde{\bm{\Lambda}}$ and returns both the Box-Cox transformed values and the optimal value for $\kappa$.

%a Nelder-Mead simplex algorithm~\cite{Nelder:65}}.

\begin{figure}[!htb]
\centering
\includegraphics[scale=0.43]{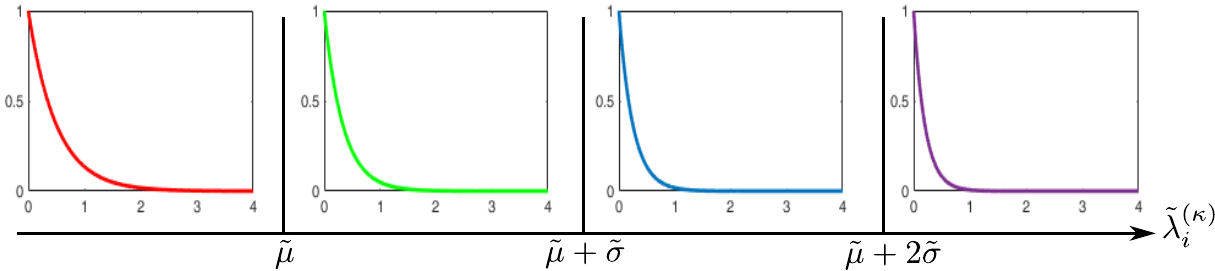}
\caption{Adaptive transfer function plots for different scaling parameters from smaller to larger (left to right).}
\label{fig:ATFplot}
\end{figure}

\medskip\noindent\textbf{Adaptive Transfer Function.}\quad Denote by $\tilde{\mu}$ and $\tilde{\sigma}$ the sample mean and sample standard deviation of the Cox-Box transformed eigenvalues $\tilde{\lambda}_{i}^{(\kappa)}$. We define an adaptive transfer function of a low-pass filter with scaling parameter $t$ as follows:
\begin{equation}
g_{t}(\tilde{\lambda}_{i}^{(\kappa)}) \!=\! \begin{cases}
e^{-\tilde{\lambda}_{i}^{\kappa}(1+2t/c)} &\!\!\!\mbox{if }  \tilde{\lambda}_{i}^{(\kappa)}<\tilde{\mu} \\
e^{-\tilde{\lambda}_{i}^{\kappa}(1+3t/c+\tilde{\sigma})} &\!\!\!\mbox{if }  \tilde{\lambda}_{i}^{(\kappa)}\in(\tilde{\mu}, \tilde{\mu}+\tilde{\sigma}) \\
e^{-\tilde{\lambda}_{i}^{\kappa}(1+4t/c+\tilde{\sigma})} &\!\!\!\mbox{if }  \tilde{\lambda}_{i}^{(\kappa)}\in(\tilde{\mu}+\tilde{\sigma}, \tilde{\mu}+2\tilde{\sigma})\\
e^{-\tilde{\lambda}_{i}^{\kappa}(1+5t/c+\tilde{\sigma})} &\!\!\!\mbox{if }  \tilde{\lambda}_{i}^{(\kappa)}> \tilde{\mu}+2\tilde{\sigma}
\end{cases}
\label{Eq:ATF}
\end{equation}
where $c=\sum_{i}\tilde{\lambda}_{i}^{(\kappa)}$ is the sum of Cox-Box transformed eigenvalues. This adaptive transfer function is illustrated in Figure~\ref{fig:ATFplot}.

The rationale behind the proposed adaptive transfer function is to create a flexible and adaptive filtering response that can better capture the characteristics of the graph data. The key properties of our adaptive transfer function include adaptability, non-uniform filtering, adaptive attenuation, and smooth transitions. These properties aim to enhance the performance of the filter in capturing the important graph features and adapting to the specific characteristics of the graph data. Specifically, the piecewise definition of the transfer function allows for non-uniform filtering behavior across different ranges of eigenvalues. By dividing the eigenvalue range into distinct cases, the transfer function can adaptively adjust its response based on the specific range in which the eigenvalue falls. This approach enables a more fine-grained control of the filtering behavior, allowing for tailored filtering characteristics to capture different graph structures or frequencies. The transfer function also adapts to the scaling parameter and adjusts the amount of attenuation applied to the eigenvalues accordingly. This adaptive attenuation ensures that the filter can better adapt to the specific characteristics of the graph data. Moreover, the presence of different cases with distinct exponential decay terms helps avoid abrupt changes in the filtering response and enables a more gradual transition between different frequency components. The smooth transitions can enhance the filtering performance by preserving important graph frequencies while suppressing unwanted noise or high-frequency components.

\medskip\noindent\textbf{Adaptive graph wavelets.}\quad Using the adaptive transfer function, we define the adaptive spectral graph wavelet basis $\bm{\Psi}_{t}$ at scale $t$ as follows:
\begin{equation}
\bm{\Psi}_{t}=\bm{\Phi}\bm{G}_{t}\bm{\Phi}^{\T},
\end{equation}
where $\bm{G}_{t}=g_{t}(\widetilde{\bm{\Lambda}}^{(\kappa)})$ and $\widetilde{\bm{\Lambda}}^{(\kappa)}$ is a diagonal matrix whose diagonal elements are the Cox-Box transformed eigenvalues. The proposed graph wavelet basis inherits many useful properties from the heat kernel, such as being multiscale and stable under perturbations of the graph. In addition, it is computationally efficient since we only need to compute the $Q$ smallest eigenvalues and associated eigenvectors of the normalized Laplacian matrix due to the rapid decay of the transfer function, where $Q\ll N$. These eigenvalues/eigenvectors can be efficiently computed using the sparse eigensolver implemented in Python.

\medskip\noindent\textbf{Learning Embeddings:}\quad Given input user embedding matrix $\bm{X}_{u}^{(\ell)}\in\mathbb{R}^{M\times F_{\ell}}$ and input item embedding matrix $\bm{Y}_{i}^{(\ell)}\in\mathbb{R}^{K\times F_{\ell}}$ of the $\ell$-th layer with $F_{\ell}$ feature maps, the output feature matrices $\bm{X}_{u}^{(\ell+1)}$ and $\bm{Y}_{i}^{(\ell+1)}$ of CAGLE are obtained by applying the following layer-wise propagation rule:
\begin{equation}
\begin{bmatrix}
\bm{X}_{u}^{(\ell+1)} \\
\bm{Y}_{i}^{(\ell+1)}
\end{bmatrix}=\sigma\left(\bm{\Psi}_{t}\widetilde{\bm{\Lambda}}\bm{H}^{(\ell)}\bm{\Psi}_{t}^{-1}
\begin{bmatrix}
\bm{X}_{u}^{(\ell)} \\
\bm{Y}_{i}^{(\ell)}
\end{bmatrix}\bm{W}^{(\ell)}\right),
\label{Eq:AGCNprop}
\end{equation}
where $\bm{W}^{(\ell)}\in\mathbb{R}^{F_{\ell}\times F_{\ell+1}}$ is a trainable weight matrix  with $F_{\ell +1}$ feature maps, $\sigma(\cdot)$ is the point-wise sigmoid activation function, and $\bm{H}^{(\ell)}$ is a diagonal matrix given by
\begin{equation}
\bm{H}^{(\ell)}=\sigma\bigl(\bm{G}_{t}\odot\bm{\Theta}^{(\ell)}\bigr),
\end{equation}
which controls how the graph frequencies (eigenvalues) are attenuated by learning a diagonal weight matrix $\bm{\Theta}^{(\ell)}$, and $\odot$ denotes the point-wise element (Hadamard) product.

The inputs of the first layer are side information matrices $\bm{X}_{u}^{(0)}\in\mathbb{R}^{M\times F}$ and $\bm{Y}_{i}^{(0)}\in\mathbb{R}^{K\times F}$, which are often randomly initialized with appropriate values generated from a zero mean Gaussian distribution in an effort to prevent the problem of exploding or vanishing gradients. For simplicity, we assume that the feature dimensions are equal for all layers, i.e., $F_{\ell}=P$ for all $\ell\ge 1$, with $P \ll\min(M,K)$.

\medskip\noindent\textbf{Concatenation.}\quad We combine all layers by concatenating their outputs as follows:
\begin{equation}
\bm{X}_{u}=\vc{\parallel}{L}{\ell=0}
\bm{X}_{u}^{(\ell)}\quad\text{and}\quad \bm{Y}_{i}=\vc{\parallel}{L}{\ell=0} \bm{Y}_{i}^{(\ell)}
\end{equation}
resulting in a user embedding matrix $\bm{X}_{u}\in\mathbb{R}^{M\times (1+L)P}$ and an item embedding matrix $\bm{Y}_{i}\in\mathbb{R}^{K\times (1+L)P}$, where $\parallel$ denotes column-wise concatenation.

\medskip\noindent\textbf{Model Prediction.}\quad The concatenated matrices $\bm{X}_{u}$ and $\bm{Y}_{i}$ of all layers' embeddings can be used as input for downstream tasks such as classification, clustering, and recommendation. Since the latter task is the focus of this paper, we apply inner product to estimate the predicted items for a target user as follows:
\begin{equation}
\widehat{\bm{R}}=\bm{X}_{u}\bm{Y}_{i}^{\T},
\end{equation}
where $\widehat{\bm{R}}=(\widehat{\bm{R}}_{ui})$ is an $M\times K$ matrix of estimated user-item preference scores. Each entry $\widehat{\bm{R}}_{ui}=\bm{x}_{u}\bm{y}_{i}^{\T}$ is the rating score of item $i$ being recommended to user $u$, where $\bm{x}_{u}$ and $\bm{y}_{i}$ denote the rows of $\bm{X}_{u}$ and $\bm{Y}_{i}$, respectively.

\medskip\noindent\textbf{Loss function.}\quad The parameters of the proposed CAGLE model are learned by minimizing a regularized Bayesian personalized ranking (BPR) loss function~\cite{Rendle:09} given by
\begin{equation}
\begin{split}
\mathcal{L} &=-\sum_{(u,i,j)\in\mathcal{D}}\log\sigma(\widehat{\bm{R}}_{ui}-\widehat{\bm{R}}_{uj})\\
&\quad +\frac{\eta}{2}(\Vert\bm{X}_{u}\Vert_{F}^{2}+\Vert\bm{Y}_{i}\Vert_{F}^{2}),
\end{split}
\end{equation}
with respect to $\bm{W}=\{\{\bm{W}^{(\ell)}\}_{\ell=0}^{L-1},\{\bm{\Theta}^{(\ell)}\}_{\ell=0}^{L-1},\bm{X}_{u}^{(0)},\bm{Y}_{i}^{(0)}\}$, a set of learnable parameters, over the pairwise training set
$$\mathcal{D}=\{(u,i,j): (u,i)\in\mathcal{U}\times\mathcal{I}_{u}^{+}\text{ and } (u,j)\in\mathcal{U}\times\mathcal{I}_{u}^{-}\},$$
where $\sigma(\cdot)$ is the sigmoid function, and $\eta$ is a tuning parameter that controls the strength of the regularization term, which is added to impose smoothness constraint on the solutions and also to prevent over-fitting.

\begin{figure}[!htb]
\centering
\includegraphics[scale=0.7]{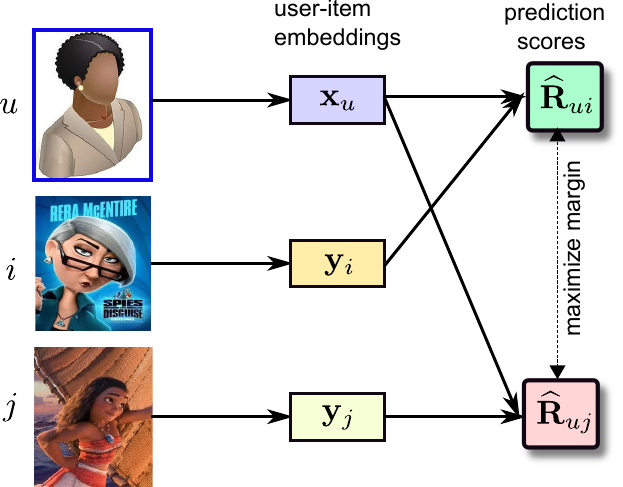}
\caption{Illustration of training triple in BPR.}
\label{fig:BPR}
\end{figure}

A training triple $(u,i,j)\in\mathcal{D}$, which consists of one user and two items, indicates that user $u$ is assumed to prefer item $i$ over item $j$. In addition, minimizing $\mathcal{L}$ is tantamount to determining whether the observed pair $(u,i)$ should have a higher user-item preference score than the unobserved pair $(u,j)$. Hence, the aim is to maximize the margin between the observed score $\widehat{\bm{R}}_{ui}$ and the unobserved score $\widehat{\bm{R}}_{uj}$, as shown in Figure~\ref{fig:BPR}.

\noindent The loss function can be rewritten as
\begin{equation}
\begin{split}
\mathcal{L} &=-\sum_{(u,i,j)\in\mathcal{D}}\log\sigma(\bm{x}_{u}\bm{y}_{i}^{\T}-\bm{x}_{u}\bm{y}_{j}^{\T}) \\
&\quad +\frac{\eta}{2}\left(\sum_{u=1}^{M}\Vert\bm{x}_{u}\Vert_{2}^{2}+\sum_{i=1}^{K}\Vert\bm{y}_{i}\Vert_{2}^{2}\right),
\end{split}
\end{equation}
\noindent Using the derivative property $\sigma'(\bm{z})=\sigma(\bm{z})(1-\sigma(\bm{z}))$ of the sigmoid function $\sigma(\cdot)$, the partial derivatives of $\mathcal{L}$ with respect to $\bm{x}_{u}$ and $\bm{y}_{i}$ are then given by
\begin{equation}
\frac{\partial\mathcal{L}}{\partial\bm{x}_{u}} =-(1-\sigma(\bm{x}_{u}(\bm{y}_{i}-\bm{y}_{j})^{\T}))(\bm{y}_{i}-\bm{y}_{j})+\eta\,\bm{x}_{u}
\end{equation}
and
\begin{equation}
\frac{\partial\mathcal{L}}{\partial\bm{y}_{i}} =-(1-\sigma(\bm{x}_{u}(\bm{y}_{i}-\bm{y}_{j})^{\T}))\bm{x}_{u}+\eta\,\bm{y}_{i}
\end{equation}
Once the learning of the model's parameters is done, we can then compute a personalized ranked list for a user $u$ based on the values of the estimated scores $\widehat{\bm{R}}_{ui}$ over all items.

\section{Experiments}
In this section, we conduct experiments to assess the performance of the proposed CAGLE framework in comparison with competing baseline models on several benchmark datasets. While presenting and analyzing our experimental results, we aim to answer the following research questions:

\begin{itemize}
\item \textbf{RQ1:} How does CAGLE perform in comparison with state-of-the-art recommendation models?
\item \textbf{RQ2:} How does recommendation performance vary across users with different levels of activity?
\item \textbf{RQ3:} How does CAGLE alleviate the cold-start recommendation problem?
\item \textbf{RQ4:} What is the effect of hyperparameters on the performance of CAGLE?
\end{itemize}

\bigskip\noindent{\textbf{Datasets.}}\quad We demonstrate and analyze the performance of the proposed approach on four benchmark datasets: MovieLens-1M~\cite{harper2015movielens},  Gowalla\footnote{https://github.com/wubinzzu/NeuRec/tree/master/dataset}, Amazon Musical Instruments and Amazon Arts, Crafts and Sewing~\cite{McAuley:19}.

\begin{itemize}
\item MovieLens-1M (\textbf{ML-1M}): This user-movie dataset consists of ratings (on a 5-star scale of 1 to 5), collected from MovieLens, a movie recommendation service that recommends movies for its users to watch using collaborative filtering. Each user has rated at least 20 movies. We discard users and items with fewer than five interactions.
\item Gowalla: This is a dataset from Gowalla, a location-based social networking website where users share their locations by checking-in. We perform data filtering by keeping users and items that have at least twenty interactions.
\item Amazon Musical Instruments (\textbf{MI}): This dataset contains data about music instruments products in Amazon. We only keep users and items with at least ten interactions.
\item Amazon Arts, Crafts and Sewing (\textbf{ACS}): This dataset contains data about Arts, Crafts and Sewing products in Amazon. Only users and items with at least twenty interactions are kept.
\end{itemize}

All explicit data are binarized and interpreted as implicit feedback. Dataset statistics are summarized in Table~\ref{Tab:data1}. We see that the Gowalla and Amazon datasets have the highest sparsity level (i.e., ratio of observed to total interactions).

\begin{table}[!htb]
\setlength{\tabcolsep}{.4em}
\caption{Summary statistics of datasets.}
\label{Tab:data1}
\centering
\medskip
\begin{tabular}{lrrrc}
\toprule
\multirow{1}[1]{*}{Dataset} & \multicolumn{1}{c}{\#Interactions} &
\multicolumn{1}{c}{\#Users} & \multicolumn{1}{c}{\#Items} &
\multicolumn{1}{c}{Sparsity (\%)} \\
\midrule
ML-1M & 1,000,209 & 6040 &3704  & 99.95\\
Gowalla & 1,027,370 &  29,858  & 40,981 & 99.99\\
MI & 1,512,530 & 112,222 & 903,330 &  99.99  \\
ACS & 2,875,917 & 302,809 & 1,579,230 & 99.99 \\
\bottomrule
\end{tabular}
\end{table}

\medskip\noindent{\textbf{Baseline Methods.}}\quad  We evaluate the performance of CAGLE against several collaborative filtering baselines, including, Neural Matrix Factorization (NeuMF)~\cite{Xiangnan:17}, Spectral Collaborative Filtering (SpectralCF)~\cite{LZheng:18}, Neural Graph Collaborative Filtering (NGCF)~\cite{XWang:19}, Light Graph Convolutional network (LightGCN)~\cite{XiangnanHe:20}, Simple Graph Contrastive Learning method (SimGCL)~\cite{yu2022graph}, Hypergraph Contrastive Collaborative Filtering (HCCF)~\cite{xia2022hypergraph}, and Deoscillated Graph Collaborative Filtering (DGCF)~\cite{liu2020deoscillated}. For baselines, we mainly consider methods that are closely related to CAGLE and/or the ones that are state-of-the-art recommendation models. A brief description of these strong baselines can be summarized as follows:

\begin{itemize}
\item \textbf{NeuMF}~\cite{Xiangnan:17} is a neural collaborative filtering approach, which combines a generalized matrix factorization model and the nonlinear multilayer perceptron network by concatenating their last hidden layers to model user-item interactions.
\item \textbf{SpectralCF}~\cite{LZheng:18} is a deep recommendation model, which performs collaborative filtering in the spectral domain of a user-item bipartite graph by leveraging a convolution filter defined in terms of the graph Fourier basis.
\item \textbf{NGCF}~\cite{XWang:19} is a graph collaborative filtering framework, which integrates the user-item interactions into the embedding process by taking into account the high-order connectivity information obtained from multiple embedding propagation layers using a variant of graph convolutional networks.
\item \textbf{LightGCN}~\cite{XiangnanHe:20} is a simplified neural graph collaborative filtering approach based on the simple graph convolution, which successively removes the nonlinear activation functions and collapses the weight matrices between consecutive layers.
\item \textbf{SimGCL}~\cite{yu2022graph} is a contrastive learning-based recommendation method that discards the graph augmentations and instead adds uniform noises to the embedding space for creating contrastive views.
\item \textbf{HCCF}~\cite{xia2022hypergraph} is a self-supervised recommendation framework that leverages a hypergraph-enhanced cross-view contrastive learning architecture. It aims to capture both local and global collaborative relations among users and items.
\item \textbf{DGCF}~\cite{liu2020deoscillated} is a recommender system model designed to address the oscillation problem by incorporating cross-hop propagation layers and locality-adaptive layers, which remedy the bipartite propagation structure and provide control over the amount of information being propagated.
\end{itemize}

\medskip\noindent{\textbf{Evaluation Metrics.}}\quad In recommender systems, the ranking of predictions matters more than an individual item's prediction score. The effectiveness of the proposed CAGLE model and baselines is assessed using two rank-based metrics, Recall$@k$ and normalized discounted cumulative gain ($\text{NDCG}@k$), by producing a $\text{top-}k$ list of items for each user based on the predicted preferences. $\text{Recall}@k$ is defined as
\begin{equation}
\text{Recall}@k = \frac{\text{\#items the user interacted with in top-}k}{\text{total \#items the user has interacted with}},
\end{equation}
which is the proportion of relevant items that are recommended in the $\text{top-}k$ list. Note that $\text{Recall}@k$ only focuses on top $k$ items in the predicted list. A larger value of $\text{Recall}@k$ indicates a better performance of the recommendation model.

On the other hand, the discounted cumulative gain $\text{DCG}@k$ is a rank-aware metric~\cite{Jarvelin:02}, which takes into consideration the order of recommended items in the list and is defined as
\begin{equation}
\text{DCG}@k =\sum_{i=1}^{k}\frac{2^{rel_{i}} -1}{\log_{2}(i+1)},
\end{equation}
where $rel_{i}$ is the relevance (binary or real number) of the item at index $i$. Since we are dealing with binary relevances, $rel_{i}$ is 1 if item at index $i$ is relevant and 0 otherwise. The normalized discounted cumulative gain $\text{NDCG}@k$ is simply defined as $\text{DCG}@k$ divided by its maximum possible value $\text{IDCG}@k$, where all the held-out items are ranked at the top of the list. Larger $\text{NDCG}@k$ values indicate better performance of the recommendation algorithm. For both metrics, we report the average scores over all test users.

\medskip\noindent{\textbf{Implementation Details.}}\quad For fair comparison, we implement the proposed model and baseline methods using NeuRec\footnote{https://github.com/wubinzzu/NeuRec}, an open-source Python library for neural recommender models, and we follow similar experimental setup to the experiments in~\cite{xue2019deep,he2018outer,chae2018cfgan,LZheng:18}. We split the datasets into 80\% training and 20\% testing. As suggested in~\cite{Bingbing:19}, we set the elements of the spectral graph wavelet matrix $\bm{\Psi}_{t}$ and its inverse $\bm{\Psi}_{t}^{-1}$ to 0 if they are smaller than 1e-7 in an effort to improve computational efficiency. For both CAGLE and SpectralCF, we set the embedding dimension to 64. We use the Glorot initialization for all models. We set the depth of the CAGLE network to three layers using the Adam optimizer with a batch size of 1024. For the number of epochs, we use an early stopping strategy~\cite{XiangnanHe:20}. The learning rate and the scale $t$ of the graph wavelet basis are chosen via grid search with cross-validation over the sets $\{0, 0.001, 0.01,\dots,0.1\}$ and $\{0.1, 0.2, 0.4,\dots,2\}$, respectively.

\subsection{Recommendation Performance}
In order to answer \textbf{RQ1}, we report the performance of CAGLE and baseline methods in Table~\ref{Tab:performance} using Recall@20 and NDCG@20 as evaluation metrics. Each metric is averaged across all test users. As can be seen in Table~\ref{Tab:performance}, the results show that the proposed CAGLE model consistently outperforms the baselines across all datasets for top-20 recommendation. Compared to the second best performing model (i.e., SimGCL), our approach yields relative improvements of 10.45\%,  5.04\%, 21.15\%, 11.86\%  in terms of Recall@20 on the ACS, ML-1M, MI and Gowalla datasets, respectively. Similarly, in terms of NDGC@20 the relative improvements of CAGLE over the best performing baseline are 8.87\%, 2.34\%, 15.24\%, 8.78\%. Notice that the highest and lowest performance improvements are achieved on the MI and ML-1M datasets, respectively. For comparing the means of CAGLE and the best performing DGCF baselines, we perform a two-sample $t$-test, where the null hypothesis is that the means are equal, with an alternative hypothesis that the means are unequal (i.e., two-sided hypothesis test). As shown in Table~\ref{Tab:performance}, all the reported $p$-values do not exceed the significance level 0.05; thereby the differences between the means are statistically significant.

\begin{figure}[!htb]
\centering
\includegraphics[scale=.62]{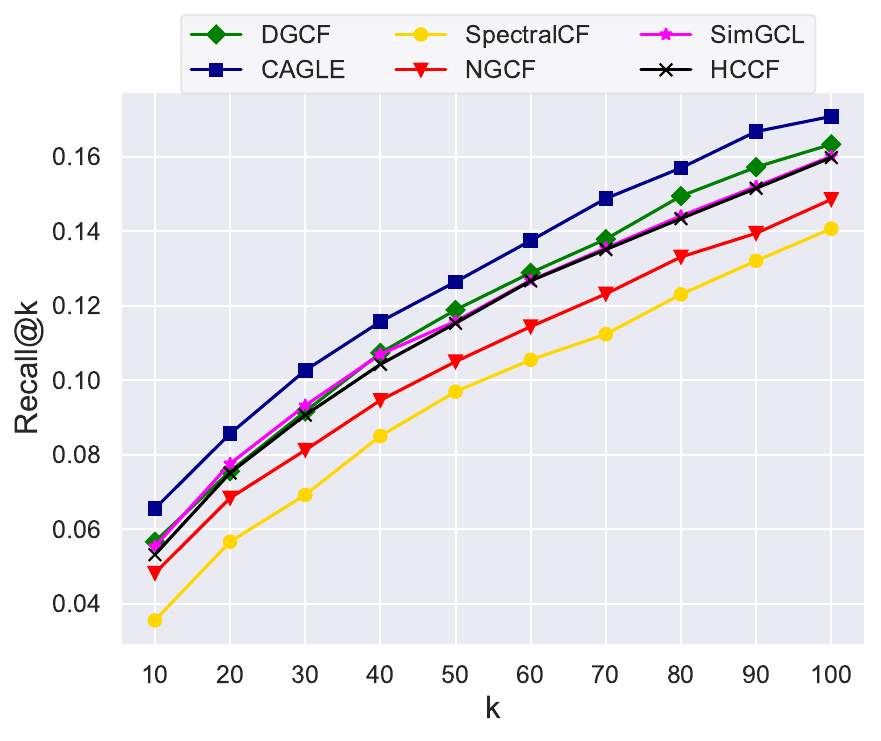}\\[1ex]
\includegraphics[scale=.62]{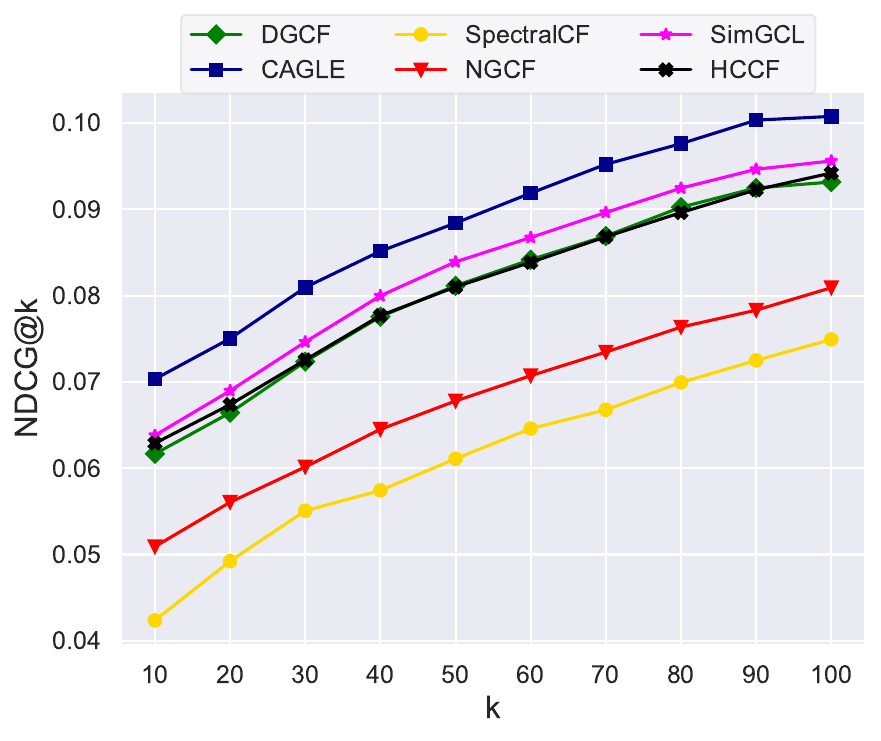}
\caption{Performance comparison of CAGLE and baselines on the ACS dataset in terms of $\text{Recall}@k$ and $\text{NDCG}@k$.}
\label{Fig:ACS-Topk}
\end{figure}

\begin{figure}[!htb]
\centering
\includegraphics[width=3.5in]{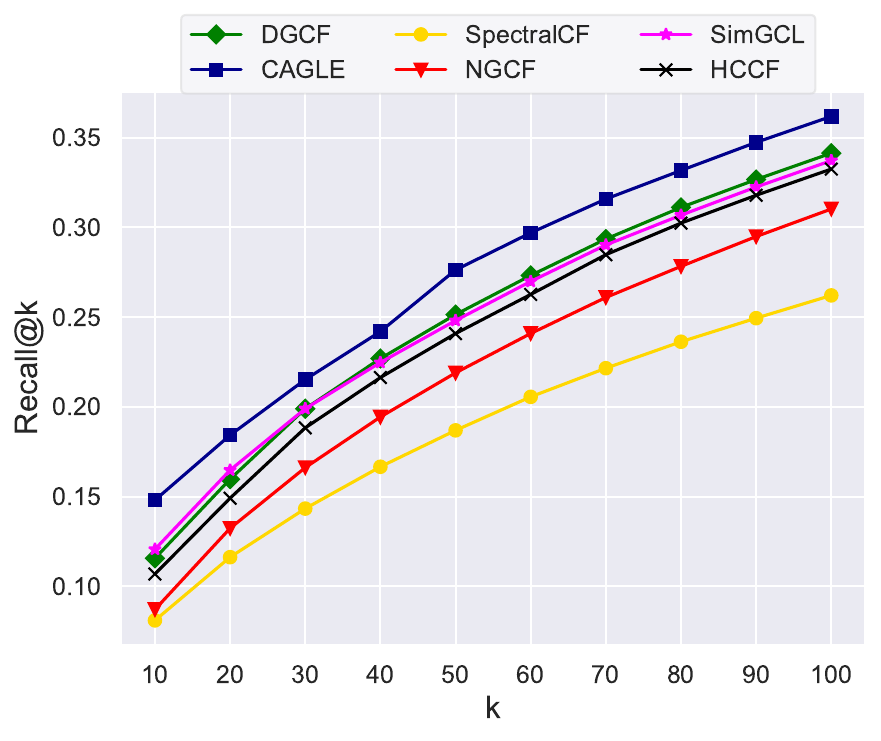}\\[1ex]
\includegraphics[width=3.5in]{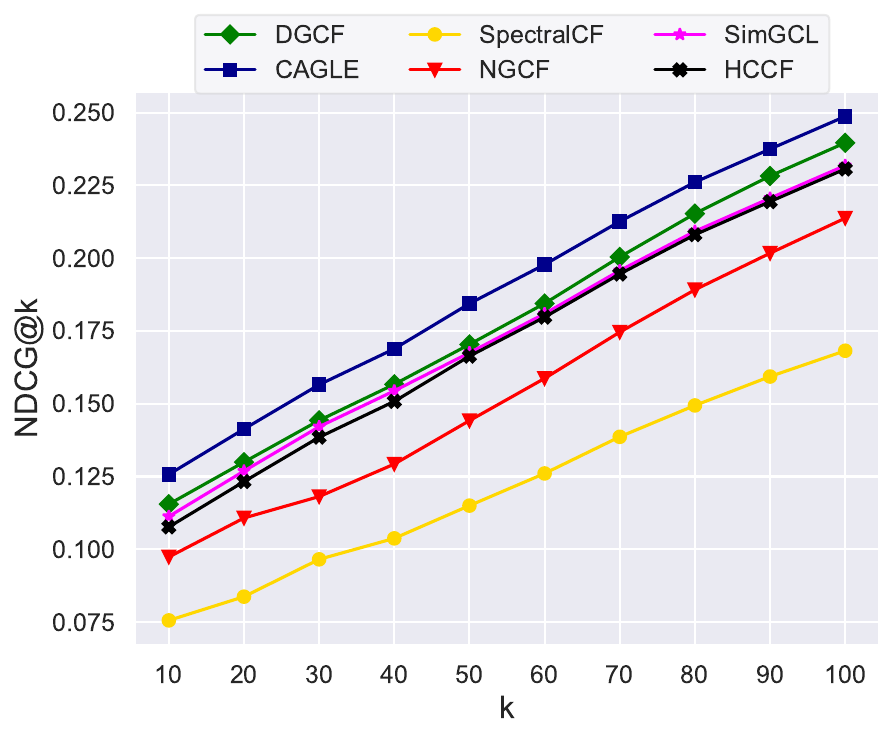}
\caption{Performance comparison of CAGLE and baselines on the Gowalla dataset in terms of $\text{Recall}@k$ and $\text{NDCG}@k$.}
\label{Fig:Gowalla-Topk}
\end{figure}

Figures~\ref{Fig:ACS-Topk}-to-\ref{Fig:Gowalla-Topk} show the performance comparison of CAGLE and baselines on all datasets in terms of $\text{Recall}@k$ and $\text{NDCG}@k$ by varying the value of $k$ for top-$k$ recommendation. As can be seen, the proposed approach yields the best recommendation performance in comparison with all baseline methods. In particular, we can observe that CAGLE performs better than DGCF and SimGCL on the large MI dataset. Moreover, CAGLE outperforms SpectralCF by a significant margin on the large MI dataset. This better performance of the proposed model is largely attributed to the fact that spectral graph wavelets yield localization of graph signals in both spatial and spectral domains, resulting in improved recommendation quality.

\begin{table*}[!htb]
\centering
\caption{Recommendation results in terms of $\text{Recall}@20$ and $\text{NDCG}@20$. The best results for each evaluation metric are highlighted in bold. For all methods, the $p$-values are less than $10^{-4}$.} %while the second best ones are underlined}.
\label{Tab:performance}
\setlength{\tabcolsep}{5.4pt}
\medskip
\begin{tabular}{lcccccccc}
\toprule[1pt]
\multirow{2}[3]{*}{Method} & \multicolumn{2}{c}{ACS} &
\multicolumn{2}{c}{ML-1M} & \multicolumn{2}{c}{MI} &
\multicolumn{2}{c}{Gowalla} \\
\cmidrule(lr){2-3} \cmidrule(lr){4-5} \cmidrule(lr){6-7} \cmidrule(lr){8-9}
  & $\text{Recall}@20$ & $\text{NDCG}@20$ & $\text{Recall}@20$ &
$\text{NDCG}@20$ & $\text{Recall}@20$ & $\text{NDCG}@20$ &
$\text{Recall}@20$ & $\text{NDCG}@20$\\
\midrule[.8pt]
NeuMF~\cite{Xiangnan:17} & 0.0502& 0.0464 &0.2262&  0.2259&0.0192&0.01641 & 0.0941&0.0751\\
SpectralCF~\cite{LZheng:18} & 0.0566 &  0.0492  &0.2427&0.2323& 0.0251&0.0192& 0.1162&0.0837\\
NGCF~\cite{XWang:19} & 0.0683& 0.0560&0.2732& 0.2435& 0.0327&0.0240 & 0.1323&0.1107\\
LightGCN~\cite{XiangnanHe:20} & 0.0721 &0.0651&0.2869&0.2521&0.0405&0.0307&0.1548&	0.1230\\
HCCF~\cite{xia2022hypergraph} & 0.0750 &0.0673&0.3115&0.2519&0.0423&0.0303&0.1491&	0.1232\\
DGCF~\cite{liu2020deoscillated} & 0.0755 &0.0664 & 0.3074&0.2549&0.0432&	0.0318& 0.1597&0.1299\\
SimGCL~\cite{yu2022graph} & 0.0775& 0.0689&0.3193&  0.2565&0.0434&0.0328 & 0.1647&0.1268\\
\midrule[.8pt]
CAGLE &\textbf{0.0856}&\textbf{0.0750}& \textbf{0.3354}& \textbf{0.2625}&\textbf{0.0526}&\textbf{0.0378} & \textbf{0.1842} &\textbf{0.1413}\\
%\midrule
%\textcolor{blue}{$p$-value} & $<$0.0001  & $<$0.0001  & $<$0.0001  & $<$0.0001  & $<$0.0001  & $<$0.0001 & $<$0.0001 & $<$0.0001 \\
\bottomrule[1pt]
\end{tabular}
\end{table*}

\begin{figure}[!htb]
\centering
\includegraphics[scale=.60]{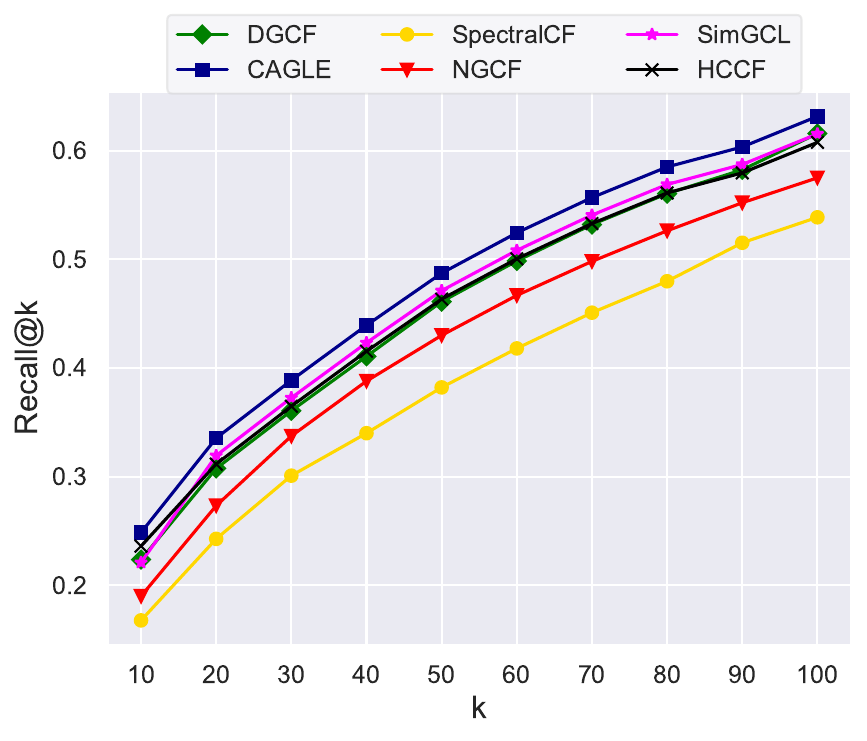}\\[1ex]
\includegraphics[scale=.60]{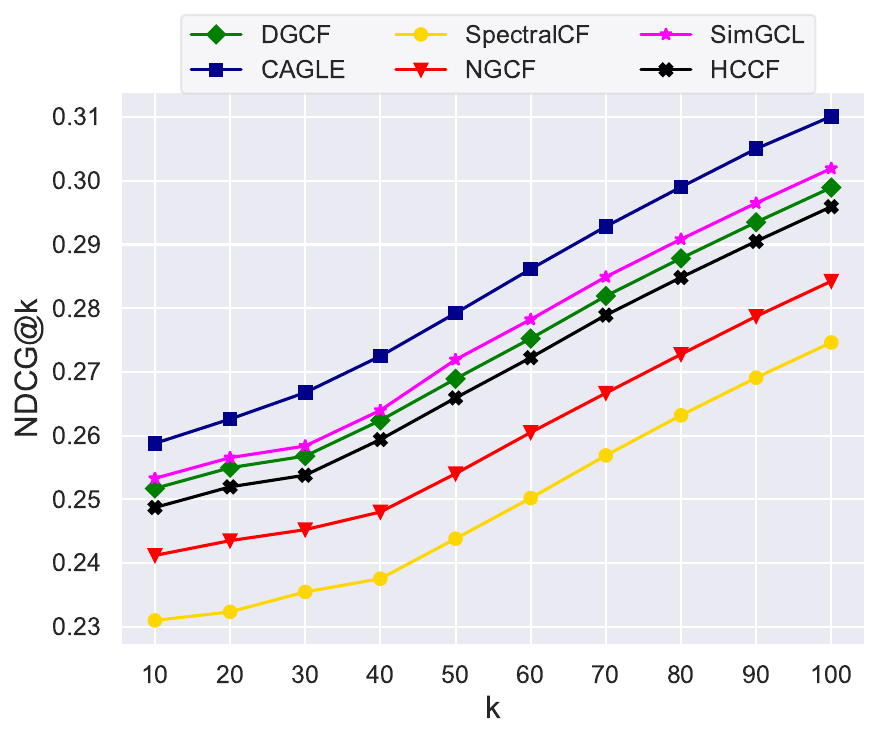}
\caption{Performance comparison of CAGLE and baselines on the ML-1M dataset in terms of $\text{Recall}@k$ and $\text{NDCG}@k$.}
\label{Fig:ML1M-Topk}
\end{figure}

\begin{figure}[!htb]
\centering
\includegraphics[scale=.60]{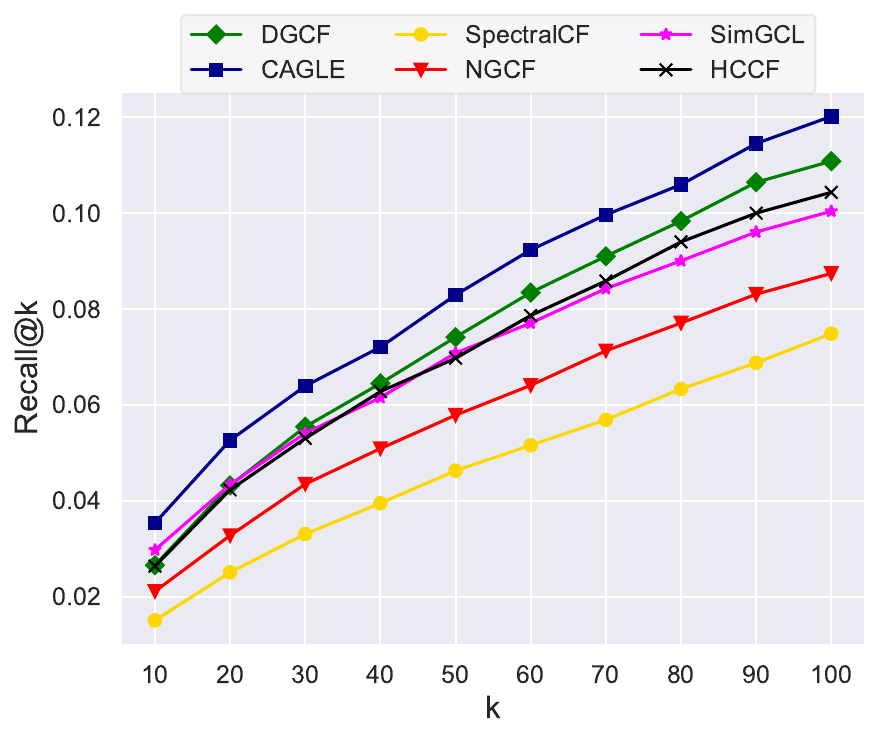}\\[1ex]
\includegraphics[scale=.60]{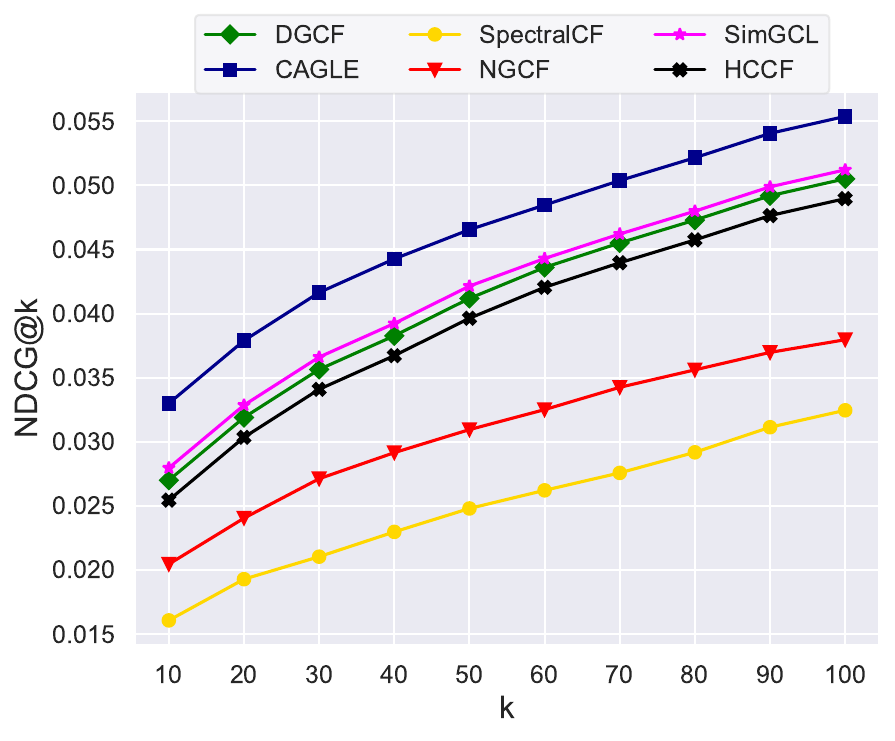}
\caption{Performance comparison of CAGLE and baselines on the MI dataset in terms of $\text{Recall}@k$ and $\text{NDCG}@k$.}
\label{Fig:MI-Topk}
\end{figure}

\subsection{Performance across Users with Different Interaction Levels}
\begin{figure}[!htb]
\centering
\includegraphics[scale=.62]{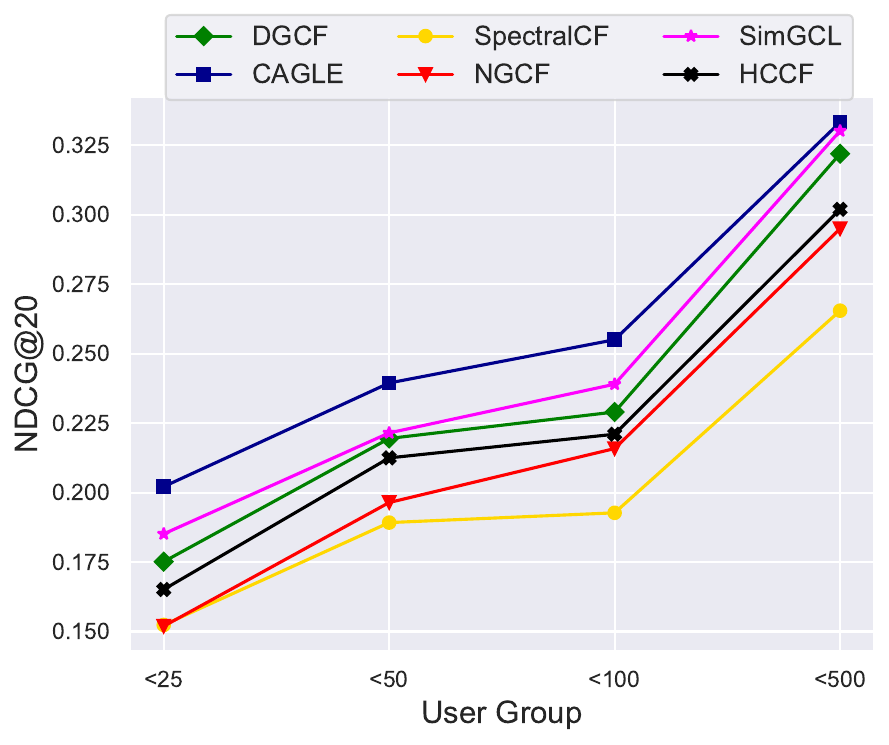}\\[2ex]
\includegraphics[scale=.62]{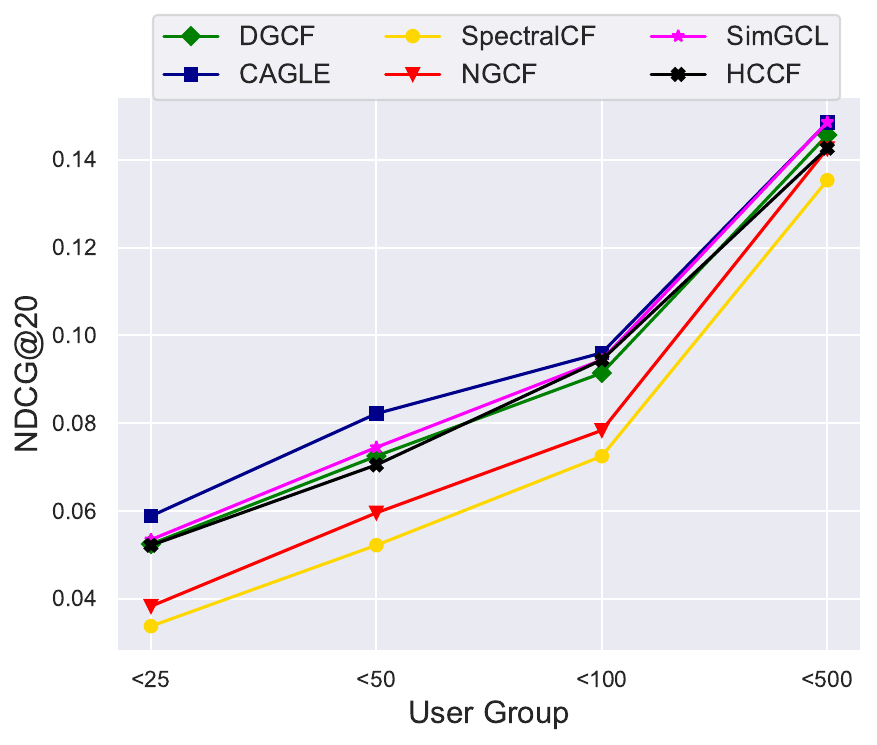}
\caption{Performance comparison across users with varying interaction levels on the ML-1M (top) and ACS (bottom) datasets.}
\label{fig:groups}
\end{figure}

To answer \textbf{RQ2}, we consider the ACS and ML-1M datasets for the sake of illustration by partitioning the test sets into four groups, each of which contains users who have interacted with a certain number of items in a specific range (e.g., the first group includes users who have interacted with less than 25 items). As shown in Figure~\ref{fig:groups}, the proposed model outperforms the baselines on all user groups. In particular, CAGLE outperforms DGCF and SimGCL on all user groups in the ML-1M dataset, indicating that our approach improves recommendation performance for users who have interacted with a relatively small number of items. The improvements are particularly noticeable on the ML-1M dataset, as the SimGCL baseline performs generally better than the other baselines when users have interacted with less than 25 items, whereas our proposed approach yields a relative improvement of approximately 7\% over these competing baselines, which is a considerable improvement within this experimental settings. A similar trend can be observed on the ACS dataset, indicating performance consistency across various datasets. This better performance is largely attributed to the fact that unlike the graph Fourier transform, graph wavelets are sparse, multiscale, and localized in the graph node domain.

\subsection{Cold-Start Recommendation Results}
Collaborative filtering inherently suffers from the data sparsity problem, which often arises when users rate or interact with a small number of items. This common issue is known as the cold-start problem, which tends to hinder the ability of a recommender system to provide good quality recommendations. To assess the performance of our proposed model in mitigating the cold-start problem (\textbf{RQ3}), we evaluate the quality of recommendations made by CAGLE for cold-start users in comparison with the recent Deoscillated Adaptive Graph Collaborative Filtering (DGCF) model~\cite{liu2020deoscillated} on the ACS dataset by creating different training sets with several degrees of sparsity. This is accomplished by varying the number of items, associated with each user, in the set $\{3, 5, 7, 9, 12\}$. The results are reported in Table~\ref{Tab:coldstart}, which shows that both CAGLE and DGCF suffer from the cold-start problem, resulting in degraded performance as the number of cold-start users decreases. However, CAGLE significantly outperforms DGCF in terms of both Recall@20 and DGCF@20. On average, CAGLE outperforms DGCF by relative improvements of 43.56\% and 47.63\% in Recall@20 and NDCG@20, respectively. Hence, CAGLE is capable of helping alleviate the cold-start problem in collaborative filtering, thanks in large part not only to its adaptive behavior, but also to its ability to localize graph signals in both spatial and spectral domains.

\begin{table}[!htb]
\caption{Performance comparison in terms of Recall@20 and NDCG@20 in the sparse training sets by varying the number of items associated with each user in the ACS dataset. The average results are reported and bold numbers indicate better performance.}
\label{Tab:coldstart}
\setlength{\tabcolsep}{2.5pt}
\centering
\medskip
\small
\begin{tabular}{llccccc}
\toprule[1pt]
 & \# Items           & 3      & 5      & 7      & 9      & 12      \\
\midrule[.8pt]
\multirow{3}{*}{Recall@20}& DGCF~\cite{liu2020deoscillated}        & 0.017  & 0.021  & 0.024  & 0.026  & 0.029  \\
% &Fixed-CAGLE & 0.009  & 0.011  & 0.014  & 0.019  & 0.023  \\
& CAGLE       & \textbf{0.025}  &\textbf{ 0.029}  &\textbf{ 0.034}  &\textbf{ 0.038}  & \textbf{0.042}  \\
\cmidrule(lr){2-7}
 & Improvement (\%) & 47.06   & 38.09 & 41.67 & 46.15 & 44.83 \\
\midrule[.8pt]
\multirow{3}{*}{NDCG@20}   & DGCF~\cite{liu2020deoscillated}        & 0.015  & 0.019  & 0.022  & 0.024  & 0.026  \\
%& Fixed-CAGLE      & 0.010  & 0.013  & 0.017  & 0.020  & 0.022  \\
& CAGLE       & \textbf{0.023}  & \textbf{0.028}  & \textbf{0.032}  & \textbf{0.035}  & \textbf{0.038}  \\

\cmidrule(lr){2-7}
& Improvement (\%) & 53.33 & 47.37 & 45.45 & 45.83 & 46.15 \\
\bottomrule[1pt]
\end{tabular}
\end{table}

\subsection{Ablation Studies}
The network depth and number of epochs play an important role in the recommendation performance of the proposed framework. We also perform an ablation experiment using a Non-Adaptive CAGLE model (CAGLE-NA for short) by removing the adaptive transfer function from CAGLE. The aim of these ablation studies is to demonstrate the effectiveness of the key components of our proposed CAGLE model.

\medskip\noindent\textbf{Mitigating the Oscillation Problem.}\quad In practice, the oscillation problem often occurs when the network depth is relatively small~\cite{liu2020deoscillated}, and refers to the trend of model performance as we vary the number of layers. To answer \textbf{RQ4}, we compare the performance of CAGLE against several strong baseline methods, including  Graph Convolutional Matrix Completion (GC-MC)~\cite{berg2017graph}, LightGCN~\cite{XiangnanHe:20}, HCCF~\cite{xia2022hypergraph}, NGCF~\cite{XWang:19}, and DGCF~\cite{liu2020deoscillated}. We also compare CAGLE to its non-adaptive counterpart (i.e., CAGLE-NA). We follow the same experimental setup as~\cite{liu2020deoscillated} by conducting experiments on ML-1M5, a filtered version of ML-1M dataset, which is obtained by retaining only ratings equal to five ~\cite{liu2020deoscillated}. The results are summarized in Table~\ref{Tab:oscillation}, which shows that CAGLE achieves better performance than the baselines across almost all layers, and performs on par with the DGCF model with four layers. Moreover, This demonstrates the ability of CAGLE to mitigate the oscillation problem. Interestingly, CAGLE-NA yields better performance than GC-MC and NGCF even without the use of adaptive mechanisms.

\begin{table}[!htb]
\caption{Performance comparison of CAGLE and baselines with increasing network depth on the ML-1M5 dataset in terms of $\text{NDCG}@20$. Boldface numbers indicate better performance.}
\label{Tab:oscillation}
\centering
\medskip
\begin{tabular}{lrrrc}
\toprule[1pt]
\multirow{1}[1]{*}{} & \multicolumn{1}{c}{Layer 1} &
\multicolumn{1}{c}{Layer 2} & \multicolumn{1}{c}{Layer 3} &
\multicolumn{1}{c}{Layer 4} \\
\midrule[.8pt]
%GCN & 0.2589& 0.2577& 0.2332& 0.2628\\
GC-MC~\cite{berg2017graph} & 0.2574& 0.2602& 0.2328& 0.2646\\
NGCF~\cite{XWang:19} & 0.2107 &0.2111 &0.1767 &0.1759  \\
LightGCN~\cite{XiangnanHe:20} & 0.2930& 0.2844& 0.2543& 0.1968 \\
HCCF~\cite{xia2022hypergraph} & 0.2948&0.2971& 0.2939& 0.2895\\
DGCF~\cite{liu2020deoscillated} & 0.3037& 0.3041& 0.3027 &0.3012  \\
\midrule[.8pt]
CAGLE-NA &0.2694 &0.2741& 0.2737& 0.2692 \\
CAGLE & \textbf{0.3086} & \textbf{0.3115} & \textbf{0.3182} & 0.3012\\
\bottomrule[1pt]
\end{tabular}
\end{table}

\smallskip\noindent We also compared CAGLE to CAGLE-NA in terms of Recall@20 and NDCG@20 in the sparse training sets by varying the number of items associated with each user in the ACS dataset. On average, CAGLE exhibits relative performance improvements that are more than double those achieved by its non-adaptive counterpart, CAGLE-NA.

\begin{figure}[!htb]
\centering
\includegraphics[scale=.6]{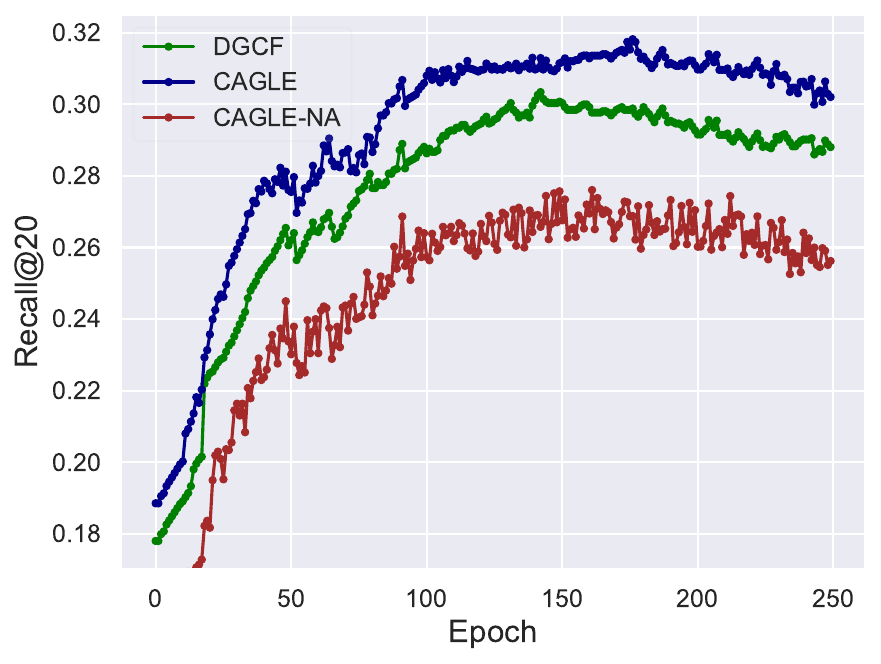}
\caption{Performance comparison of CAGLE, CAGLE-NA and DGCF with increasing number of epochs on the ML-1M5 dataset.}
\label{fig:epoch}
\end{figure}

\medskip\noindent\textbf{Effect of Number of Epochs.}\quad Figure ~\ref{fig:epoch} shows the performance comparison of CAGLE, DGCF, and CAGLE-NA in terms of Recall@20 on the ML-1M5 dataset by varying the number of epochs. Notice that CAGLE outperforms DGCF when the number of epochs exceeds 30, and yields better performance than CAGLE-NA for all epochs. Overall, CAGLE yields better performance than DGCF and CAGLE-NA, indicating its ability to learn more discriminative embeddings of users and items.

\subsection{Statistical Significance Analysis}
\begin{figure}[!htb]
\centering
\includegraphics[scale=.4]{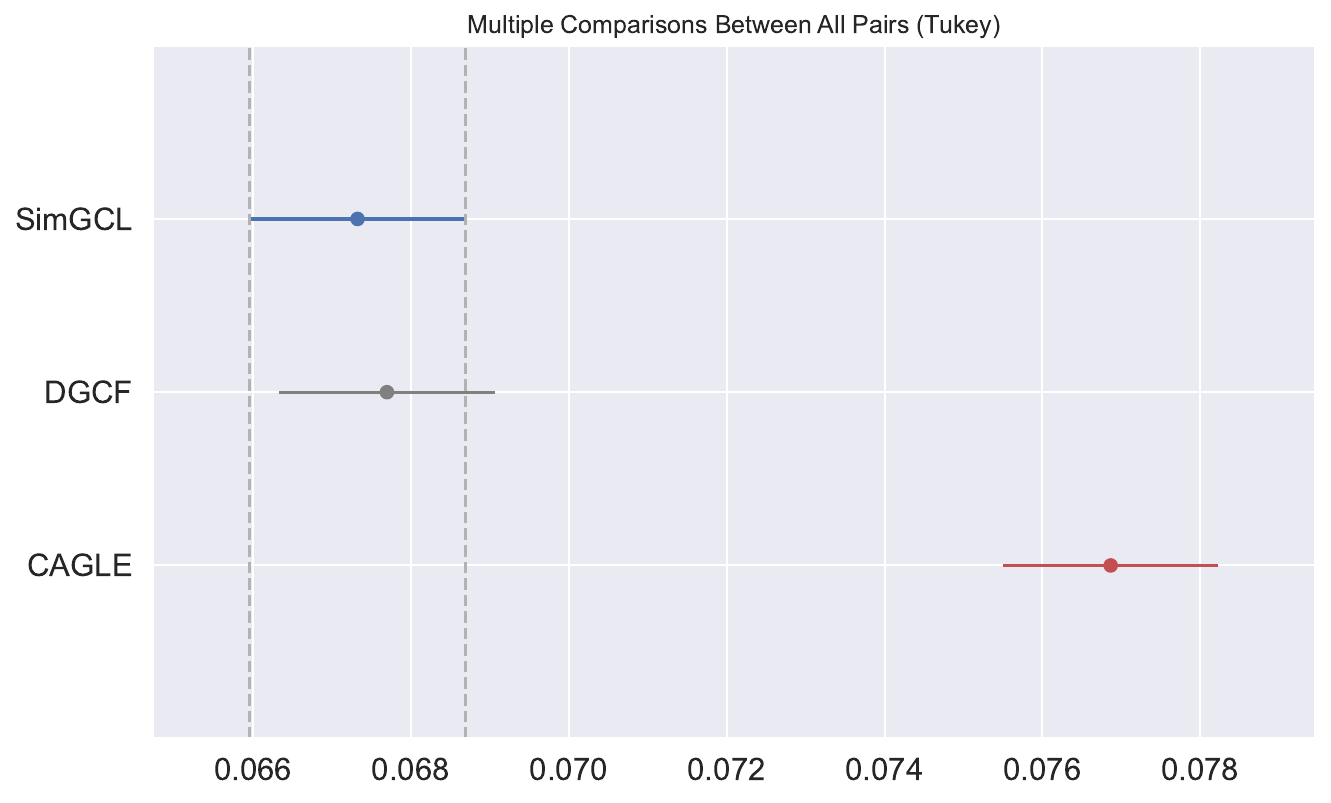}\\[2ex]
\includegraphics[scale=.4]{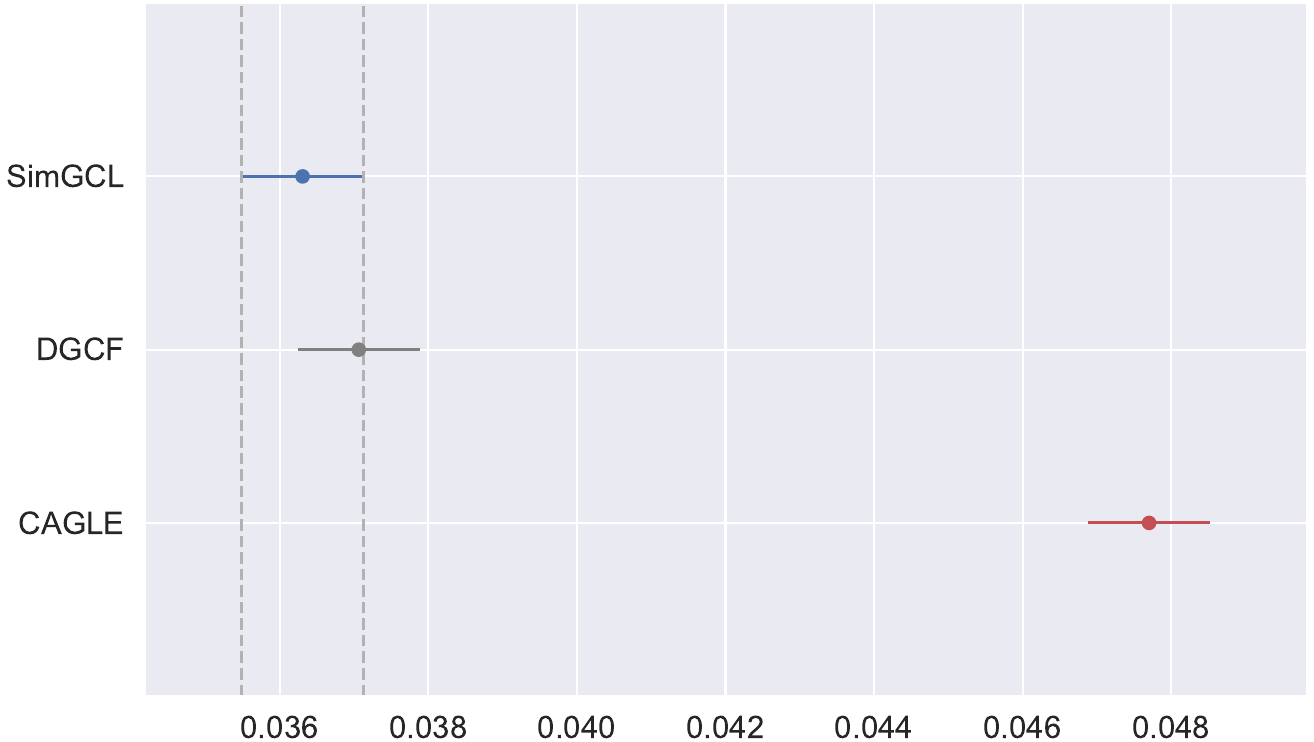}
\caption{Pairwise multiple comparison between CAGLE, DGCF and SimGCL in terms of Recall@20 using Tukey's test on the ACS dataset (top) and MI dataset (bottom).}
\label{fig:TukeyTest}
\end{figure}

We perform statistical significance tests to compare CAGLE, DGCF, and SimGCL by employing multiple pairwise comparison analysis using Tukey's test. This test is employed to compare all possible pairs of means. The results of these comparisons are depicted in Figure~\ref{fig:TukeyTest}, where 95\% confidence interval plots are represented by horizontal lines. The red line signifies the comparison interval for the mean of CAGLE. This interval does not overlap with the comparison intervals for the means of DGCF and SimGCL, which are highlighted in grey and blue colors, respectively. As indicated by the horizontal dotted lines, DGCF and SimGCL have overlapping intervals, suggesting that there is no significant difference between their group means. Conversely, CAGLE's interval does not overlap with those of SimGCL and DGCF, which implies a significant difference between the group means of CAGLE and the other two methods. Consistent with visual observations, Table~\ref{Tab:performance} corroborates these findings.

\section{Conclusion}
In this paper, we proposed a novel framework, named CAGLE, for collaborative filtering with wavelets on graph-structured data. Our approach efficiently learns low-dimensional representations of users and items in the context of spectral graph theory. We first designed an adaptive graph convolutional filter using spectral graph wavelets. Then, we introduced a deep recommendation model, which yields localization of graph signals in both spatial and spectral domains. The proposed convolutional filter is multiscale and hence can capture both local and global information from different graph neighborhoods. Our experimental results on several real-world datasets demonstrated the effectiveness of the proposed model, showing performance gains over various baseline methods. For future work, we plan to incorporate high-order neighborhood information into the proposed framework.

\bigskip\noindent\textbf{Acknowledgments.}\quad This work was supported in part by the Discovery Grants program of Natural Sciences and Engineering Research Council of Canada.

%\subsection*{Compliance with Ethical Standards}
%\noindent{\textbf{Ethical Approval}}\quad This article does not contain any studies with human participants performed by any of the authors.
%
%\smallskip\noindent{\textbf{Competing interests}}\quad The authors declare that they have no financial or personal interests to disclose.
%
%%\smallskip\noindent{\textbf{Funding}}\quad This work was supported in part by Natural Sciences and Engineering Research Council of Canada under a Discovery Grant Number N00929.
%
%\smallskip\noindent{\textbf{Availability of data and materials}}\quad The datasets used in the experiments are publicly available.

\bibliographystyle{ieeetr}
\bibliography{references} %List of references

\end{document}